\documentclass[showkeys,superscriptaddress,twocolumn,pre,showpacs,preprintnumbers]{revtex4-1}
\usepackage{amsmath,amssymb}
\usepackage{epsfig,graphics,color,calc,graphicx}
\usepackage{mathrsfs}

\newcommand{\rr}[1]{{\normalfont\textrm{#1}}}

\newcommand{\bb}[1]{{\mathbb{#1}}}

\newcommand{\plow}{p}

\newlength{\pecettawidth}
\begin{document}
\title{Effect of intracellular diffusion on current--voltage 
                 curves in potassium channels}

\author{Daniele Andreucci}
\email{daniele.andreucci@sbai.uniroma1.it}
\affiliation{Dipartimento di Scienze di Base e Applicate per 
             l'Ingegneria, Sapienza Universit\`a di Roma,
             via A.\ Scarpa 16, I--00161, Roma, Italy.}

\author{Dario Bellaveglia}
\email{dario.bellaveglia@sbai.uniroma1.it}
\affiliation{Dipartimento di Scienze di Base e Applicate per 
             l'Ingegneria, Sapienza Universit\`a di Roma,
             via A.\ Scarpa 16, I--00161, Roma, Italy.}
\affiliation{Dipartimento di Ingegneria Strutturale e Geotecnica
             Sapienza Universit\`a di Roma,
             via Eudossiana 18, I--00184 Roma, Italy.}

\author{Emilio N.M.\ Cirillo}
\email{emilio.cirillo@uniroma1.it}
\affiliation{Dipartimento di Scienze di Base e Applicate per 
             l'Ingegneria, Sapienza Universit\`a di Roma,
             via A.\ Scarpa 16, I--00161, Roma, Italy.}

\author{Silvia Marconi}
\email{silvia.marconi@sbai.uniroma1.it}
\affiliation{Dipartimento di Strutture,
             Universit\`a degli Studi Roma Tre, 
             via Corrado Segre 4, 00146 Roma, Italy.}

\begin{abstract}
We study the effect of intracellular ion diffusion on ionic currents
permeating through the cell membrane. Ion flux across the cell membrane
is mediated by special proteins forming specific channels. The structure
of potassium channels have been widely studied in recent years with
remarkable results: very precise measurements of the true current across
a single channel are now available. Nevertheless, a complete
understanding of the behavior of the channel is still lacking, though
molecular dynamics and kinetic models have provided partial insights. In
this paper we demonstrate, by analyzing the KcsA current-­voltage
currents via a suitable lattice model, that intracellular diffusion plays
a crucial role in the permeation phenomenon. The interplay between the
selectivity filter behavior and the ion diffusion in the intracellular
side allows a full explanation of the current-­voltage curves.
\end{abstract}

\pacs{87.10.-e; 87.16.-b; 87.16.Vy; 66.10.-x}

\keywords{potassium channel, KcsA, current--voltage curves, selection, 
          diffusion}



\maketitle

\section{Introduction}
\label{s:introduzione}
\par\noindent
Potassium currents across nerve membranes have been
widely studied (see, e.g., \cite{HH,NS,Hille}
and  the reviews~\cite{GBOZ,VanDongen02,Miller,FH,RCM}).
Ionic channels selecting potassium currents 
are present in almost all types of cells in all organisms
and they play many important and different functional roles.

Different types of measurements \cite{ADSEGHTM,ZM} provide
a very detailed description of the behavior of potassium channels.
In general, 
less is known on their structure~\cite{Miller}.
KcsA, a potassium channel from \textit{Streptomyces lividans}, 
is the first ion channel whose structure has been identified via 
X--ray crystallography \cite{doyle}.

All ionic channels
form selective pores in the cell membrane which open 
and close and, when in the open state, 
allow permeation of a selected ionic species (potassium in 
K$^+$--channels). Their ability to open and close, i.e., \textit{gating},
and 
their ability to allow the flux of a particular ionic 
species, i.e., \textit{selectivity}, 
are not yet completely understood. 
A lot is known in the case of the KcsA channel and, with some care, 
can be extrapolated to the whole family of K$^+$ channels. 

In KcsA, see for instance the detailed description in \cite{Miller,IOTS}, 
gating is realized via four crossing transmembrane helices on 
the intracellular side. When in the open state, a spherical water--filled 
cavity of diameter about 
10~\AA~  widens on the intracellular side of the channel up 
to the membrane plane. There a 10--15~\AA~ long and 3~\AA~ large channel  
containing the selectivity filter 
connects the cavity to the extracellular side. 

When the channel flips to the open state a solution with the 
cytoplasm concentration reaches the entrance of the channel.
Part of the ions permeates through the channel leaving an ion depleted 
region close to the pore. The typical time needed to restore the 
intracellular concentrations close to the pore will 
depend strongly on the diffusion process of ions inside the cell. 
We can then imagine that the current flowing through 
the channel will depend both on the diffusion of ions in the cytoplasm and 
on the behavior of the selectivity filter. 

The problem of computing the permeation current in the open state, 
namely, the so called \textit{true current},
has been approached theoretically by a 
large variety of methods. 
Molecular dynamics studies \cite{Aqvist,BR}
give a very detailed description,
but they usually do not provide 
macroscopic currents estimates due to the too small involved time 
scale. 
Kinetic models
\cite{Miller02,Nelson,MP,Nelson02,Nelson03,MP,VanDongen01} 
give very useful information,
since electro--physiological experiments are performed over time scales 
much longer than the atomic one, with the 
drawback of the extreme simplification on the structure of the channel.

In the recent literature it has been examined the possibility to validate 
these kinetic models by comparing the predicted behaviors with those observed 
experimentally. In particular the models have been tested against their 
ability to predict the current--voltage behavior. In this respect
very accurate experimental results have been 
published for different potassium channels, see 
\cite{MHM}, \cite{SH}, and \cite{HmK}
for the KcsA, MaxiK, and the Shaker 
channels, respectively.

Models such as those quoted above 
describe to some extent 
the dynamics of ion permeation through the selectivity filter, 
the concentration of the ion in 
the cell is introduced in the model as a constant parameter. 
In other words in those studies the diffusion of the ions inside
the cell is not taken into account. 
Our opinion is that diffusion, as explained above, must take 
an important part in the permeation phenomenon. 
In \cite{ABCM01}, inspired by \cite{VanDongen01}, we introduced 
a model where the channel is lumped to a two state stochastic point system
and the interaction between the dynamics of the 
ions inside the cell (diffusion) 
and that of the selectivity filter itself is 
taken into account.
The channel is then thought of as part of the cell more than 
as an isolated structure. 
In that paper both an analytical and Monte Carlo study showed 
the possibility to achieve gating via selection.

In this paper we examine the possibility to predict the 
behavior of the current--voltage curves (graph of the 
permeation current vs.\ the external potential difference 
applied to the membrane) on the basis of a
similar model. A modification is needed to take into account 
the effect of an external voltage difference through 
the cell membrane.  
The model is thus defined to mimic the three effects that seem to be 
the most relevant in the process:
(i) diffusion of the ions inside the cell; 
(ii) dynamics of the selectivity filter; 
(iii) dynamics of the ions inside the selectivity filter. 
We compare the current--voltage behavior predicted by our 
model with experimental results from \cite{MHM,SH,HmK}
and find a very good agreement both for the dependence of the current 
on the ion concentration and on the external voltage. 

Ion diffusion (item (i)) inside the cell is modeled 
as a symmetric random walk on a finite line. We use a one 
dimensional system in order to compute all the interesting 
quantities explicitly \cite[Appendix~B]{ABCM01}. As we shall 
comment later the dimensionality of the system affects only our 
estimate of the diffusion coefficients of ions inside the 
cytoplasm. Corrections will be introduced to compute the 
three--dimensional value of the diffusion coefficient. 

We remark that point (i) is the real distinguishing feature of our 
model. As we shall discuss later, the introduction 
of ion diffusion in the intracellular region and hence of the 
depletion phenomenon, will allow 
a full description of the permeation current behavior with respect to 
both external voltage and ion intracellular concentration.

One of the two boundary points of the finite line 
where ions diffuse is reflecting, 
whereas the other mimics the selectivity filter. 
The dynamics of the pore (item (ii)) 
is assumed to be stochastic. The pore jumps randomly between 
two states, the {\em low} and the {\em high--affinity} one. 
The dynamics of the pore is independent from that of the ions inside 
the cell. 
The chances that an ion has to enter the pore 
depend both on the ion species and on the pore state; in this way 
selection is implemented in the model. 

This description is quite faithful to the real behavior of the selectivity
filter. Indeed, two possible states are possible for the 
filter \cite{RMFMPEAFGMG}, the low and the high--affinity one. 
When the filter is in the low--affinity state permeation is favored, but
no ionic species is preferred.
Thus, in order to realize selectivity
in an efficient way, the filter has to jump between these two states. 
To our knowledge the time fraction spent by the filter in the 
low affinity--state is not known experimentally; the value predicted 
by analyzing the current--voltage curves via our model depends 
on the ionic concentration in cytosol and is of order 
$10^{-3}$. This result seems to be coherent with the qualitative 
description given in \cite[Figure~5]{VanDongen01}.

The dynamics of the ions inside the filter (item (iii)) is not modeled in 
detail, we just assume that a particle inside the pore can either 
exit the system or reenter the cell with a fixed probability. 
This ejection probability is chosen as a function of the 
voltage difference across the cell membrane. 
Our modeling of the dynamics inside the filter is, thus, reduced 
to the choice of this function. 
It is worth noting that 
the model is able to reproduce accurately 
the experimental results 
with different choices of this function, that is to say
with different descriptions of the dynamics of the ions inside the filter. 
More precisely we shall see that it is possible 
to explain the experimental results via different models for the dynamics 
of the ions inside the filter; when a different functional behavior for 
the ejection probability in terms of the external voltage is chosen, 
a different value of the typical time spent by the filter in the 
low--affinity state is found. In other words different models 
of the dynamics of the ions inside the filter are allowed provided the 
low--affinity state probability is changed suitably.

Indeed, we shall find very good results either by choosing the ejection 
probability 
function according to very well known and studied theories \cite{Nelson02}
or 
by assuming that the ejection probability for a potassium 
ion trapped inside the filter is a power law function 
of the applied (suitably renormalized) voltage difference.
In both cases our model will 
predict curves fitting cleanly the experimental result for the 
``true" ion current with reasonable values of the 
intracellular diffusion coefficient; 
but different values of the low--affinity state 
probability will be predicted.
A more precise knowledge of this probability would enable us to 
discriminate among different mechanisms.

Moreover, we remark that our model will be also able to 
explain the behavior of ionic species different from potassium  
by choosing properly the physical parameters 
appearing in the ejection probability function of the 
external voltage difference. 

We finally note that
some care has to be used when the theoretical results are compared 
with experiments. 
In the biological literature, see for instance \cite{ADSEGHTM}, 
two different types of current are reported, the \textit{apparent} 
and the \textit{true} one. 
This is connected with the peculiar behavior of ionic channels: 
two different states are observed, \textit{closed} and \textit{open} 
channel. In the first state a zero current is observed, in the second one 
the outgoing current fluctuates randomly on a very short time--scale 
and a not zero average current, called \textit{true current}, is measured. 
The true current can be accessed experimentally only if the time 
resolution of the instruments is good enough to distinguish 
neatly between the open and the closed time intervals. 
When this is not the case a different current, called 
\textit{apparent current}, is measured. This current is smaller than the 
true one since the instruments average the current also on 
time intervals in which the channel is closed. 

The model we propose here is intended to mimic the behavior of the 
ionic channel in the open state. The average outgoing flux that 
will be computed has then to be thought as a prediction for the true current. 

In Section~\ref{s:modello} we introduce the lattice model. 
Section~\ref{s:sperimentali} is devoted to the analysis of the 
experimental data. 
Our results will be discussed in Section~\ref{s:discussione}, where,
in particular, we shall comment on the physical meaning of the 
fitting parameters introduced previously.
Our conclusions are briefly summarized in 
Section~\ref{s:conclusioni}.

\section{The model}
\label{s:modello}
\par\noindent
The intracellular region is modeled via a 
finite one--dimensional 
lattice $\Lambda\subset\bb{Z}$ with $L$ sites.
Two sites of $\bb{Z}$ are said to be \textit{nearest--neighbors} 
if and only if their mutual Euclidean distance is equal to one.
The boundary $\partial\Lambda$ of $\Lambda$ is the collection of 
the two sites of $\bb{Z}$ not belonging to $\Lambda$ and neighboring 
one of the site of $\Lambda$. 
One of the two points of the boundary of $\Lambda$ is called \textit{pore}, 
and denoted by $P$. 

One ionic species performs 
an independent symmetric random walk on the lattice with reflectivity 
conditions on the site of $\partial\Lambda$ different from the pore. 
The particles on the site neighboring the pore behave in a special way 
that will be described below; 
due to the peculiarity of such a rule 
the walkers will turn out to be not independent.
The number of walkers is denoted by $N$.

The fact that the  walkers are independent on $\Lambda$ 
means that the position 
of a particle does not affect the motion of the others, in particular 
no constraint to the number of particles on each site is prescribed.
The fact that the random walks are symmetric means that 
each jump between two neighboring sites of $\Lambda$ is performed 
with probability $1/2$. Since we assumed that the boundary 
is reflecting, particles in the site neighboring 
$\partial\Lambda\setminus\{P\}$ can stay in the same site with probability 
$1/2$.

\begin{figure*}
\begin{picture}(200,200)
\put(-170,-340)
{
\includegraphics[height=24.cm,angle=0]{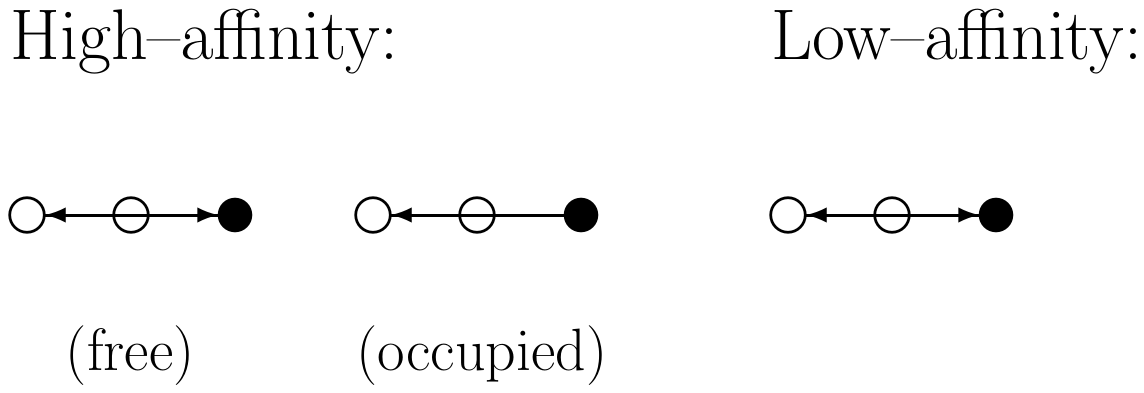}
}
\end{picture}  
\vskip 1. cm
\caption{Rule for the site close to the pore.
The black solid circle denotes the pore, 
the open circles denote sites of the lattice, 
the black arrows denotes jumps 
that are performed with probability $1/2$, the 
black lines denote jumps that cannot be performed (probability zero). 
On the left the behavior of an ion close to the 
pore in the high--affinity state is depicted:
if the pore is free, the particle jumps with 
uniform probability $1/2$ to one of the $2$ nearest neighboring sites;
if the pore is occupied, the particle jumps with 
uniform probability $1/2$ to the nearest neighboring site in 
the lattice (it cannot enter the pore)
and with probability $1/2$ the particle does not leave the site.
On the right the behavior of an ion close to the 
pore in the low--affinity state is depicted:
the particle jumps with 
uniform probability $1/2$ to one of the $2$ nearest neighboring sites;
If the pore is in the low--affinity state, the ions behave with the 
same rule as that for an ion 
faced to the free pore in the high--affinity state.
It is worth remarking that the pore in the low--affinity state cannot be 
occupied.
}
\label{f:reticolo}
\end{figure*}

Two states are allowed for the pore: \textit{high--affinity} 
and \textit{low--affinity}. The pore switches between the 
two states randomly; the probability that the pore is 
in the low--affinity state is denoted by $\plow\in[0,1]$.
Moreover the pore can be either \textit{free} or \textit{occupied} 
by an ion. 
The behavior of the particles on the site $Q$ neighboring the pore depends 
on the state of the pore itself as it is precisely stated in 
figure~\ref{f:reticolo}.
The idea is the following.
If the pore is occupied by a particle, no other 
ion can enter it. 
If the pore is (free) in the low--affinity state, 
particles can enter it; when they enter the pore 
they immediately dissociate so that, with probability $v\in[0,1]$ they 
exit the system while they reenter $\Lambda$
with probability $1-v$.
If the pore is free and in the high--affinity state, one particle
can enter it, but, once entered, 
it remains there until the pore state changes to the low--affinity one.
When this happens the ion dissociates with the same 
rule described above. 
As noted above, due to the pore rule, the walkers are not independent.

Whenever an ion exits the system another particle is put 
at random with uniform probability $1/L$ on one of the $L$ sites 
in $\Lambda$ so that the number of ions 
is kept constant.

The model described above is implemented with the 
Markov Chain described in 
detail in the appendix~\ref{s:bold}. 
An \textit{iteration} or \textit{sweep} of the chain is the 
collection of the steps that are performed at each time $n$.

As we have explained in the Introduction, the aim of this paper 
is that of computing the outgoing ionic current. This quantity 
is related to the number of particles that exit the system. 
We let $M(n)$ be the number of
particles which exited the system in the time interval $[0,n]$.
Moreover, we let the \textit{flux} 
at time $n$ be  
$M(n)/n$.
Since we defined the stochastic process in such a way that 
the number of ions keeps constant in the 
volume $\Lambda$, 
$M(n)/n$ 
approaches a constant value $f$ for $n\to\infty$.
This quantity, that we shall call the \textit{outgoing flux},
is expected to be proportional to the ``real" current 
measured experimentally. 

The existence of this limiting flux can be deduced by 
remarking that the chain is irreducible and that the space 
state is finite. So we have that there exists a stationary measure 
for the process and that, in the limit $n\to\infty$, 
the time dependent quantities 
tend to the corresponding quantities averaged against 
the stationary measure.

By exploiting one--dimensionality, the model
can be solved analytically and the outgoing flux can be computed explicitly.
In our model the 
pore is modeled in a very simple fashion, 
indeed it is just a two state Bernoulli process;
the main difficulty is, obviously, the interaction between 
the random walk inside the volume $\Lambda$ and the pore 
itself. We consider the 
stationary state reached by a walker and denote by 
$q$ the probability for the ion to occupy the 
site $Q$ of $\Lambda$ neighboring the pore.

Particles that enter the pore in the low--affinity state can exit 
the system with probability $v$. With the same probability a particle
trapped in the pore in the high--affinity state 
can exit $\Lambda$ when the pore switches to the 
low--affinity state. 
We can write the outgoing flux as
\begin{equation}
\label{flussok}
f
=
\frac{1}{2}Nq\plow v
+
\plow rv
\end{equation}
where we have denoted by $r$ the probability that in the stationary state 
the pore is occupied by a particle when it switches from the high to the 
low--affinity state. 

This one dimensional model can be solved following the 
same idea used in \cite{ABCM01}. 
We find 
\begin{equation}
\label{uk04new}
 r=\frac{N+B+C-\sqrt{(N+B+C)^2-4NC}}{2C}
\end{equation}
and
\begin{equation}
\label{uk05}
 q=\frac{2\plow}{(1-\plow)(1-r)}\frac{r}{N}
\end{equation}
where we have set 
\begin{displaymath}
A=\frac{2L^2-3L+1}{3},\;\;
B=\frac{2\plow}{1-\plow}\Big(L+\frac{v}{2}A\plow\Big)
\end{displaymath}
and
\begin{displaymath}
C=1+\plow Av
\end{displaymath}

\section{Comparison with experimental results}
\label{s:sperimentali}
\par\noindent
In \cite{MHM} it has been demonstrated that the 
KcsA, a bacterial ion channel of $K_\rr{ir}$ topology 
and very well known structure (see, for instance, \cite{Miller}
and references therein), is a potassium channel. 
In that paper very precise measurements of potassium (and some other ion 
species) currents are reported. In this section we try to explain their 
results via our model.  

Since the structure of the KcsA channel is well know, 
in this section will be mainly concerned with the 
experimental measures in \cite{MHM}. At the end a brief
analysis of the results in 
\cite{SH} and \cite{HmK}, concerning respectively the 
MaxiK and the Shaker channel, will be given.
Note that the Shaker is a voltage--activated channel of $K_\rr{v}$ topology.

\subsection{Potassium current--voltage curves}
\label{s:potassio}
\par\noindent
In \cite[Figure~2B]{MHM} the potassium current--voltage curves 
at different concentrations 
$20,50,100,200,400,800$~mM are shown.
We text the reliability of our lattice model by exhibiting neat 
fitting of those experimental data. 

\begin{table*}
\begin{center}
\begin{tabular}{c|c|c|c|c|c|c||c|c|c|c|c||c|c|c|}
\cline{2-15}
& 
\multicolumn{6}{|c||}{KcsA}
& 
\multicolumn{5}{|c||}{Shaker}
&
\multicolumn{3}{|c|}{MaxiK}
\\
\hline
\multicolumn{1}{|c|}{$N$} 
 & $800$ & $400$ & $200$ & $100$ & $50$ & $20$ 
 & $605$ & $325$ & $206$ & $73$ & $43$ 
 & $400$ & $150$ & $50$ 
\\
\hline
\multicolumn{1}{|c|}{$S_I$ ($10^6$~pA)} 
 &$3.83$ &$4.91$ &$6.33$ &$8.07$ &$12.0$ &$16.7$ 
 &$1.00$ &$1.03$ &$1.31$ &$2.09$ &$3.03$ 
 &$7.56$ &$7.03$ &$8.88$ 
\\
\hline
\multicolumn{1}{|c|}{$p$ ($10^{-3}$)} 
 &$3.93$ &$2.07$ &$1.12$ &$0.61$  &$0.32$ &$0.14$ 
 &$5.69$ &$3.91$ &$2.48$ &$1.27$ &$0.46$ 
 &$0.30$ &$0.16$ &$0.08$ 
\\
\hline
\hline
\multicolumn{1}{|c|}{$s$ ($10^{-14}$~sec)} 
 &$4.17$ &$3.25$ &$2.53$ &$1.98$  &$1.34$ &$0.96$ 
 &$16.0$ &$15.5$ &$12.2$ &$7.66$ &$5.28$ 
 &$2.12$ &$2.77$ &$1.80$ 
\\
\hline
\multicolumn{1}{|c|}{$D$ ($10^{-2}$~cm$^{2}$/sec)} 
 &$3.62$ &$4.64$ &$5.98$ &$7.62$  &$11.3$ &$15.8$ 
 &$0.88$ &$0.91$ &$1.15$ &$1.83$ &$2.65$ 
 &$7.11$ &$6.60$ &$8.34$ 
\\
\hline
\end{tabular}
\end{center}
\caption{The parameters $S_I$ and $p$ have been  
measured by fitting the experimental data in 
\cite{MHM}, \cite{HmK} (data extracted from figure~1B therein), and \cite{SH}
via (\ref{iva}) and (\ref{nelson}). 
By fitting the highest concentration set of data 
we also found 
$S_V=1.2\times 10^{-3}$, $\delta=0.19$, $l=5.5$~nm,
and $L=1.00\times 10^4$ for the KcsA,
$S_V=5.45 \times 10^{-3}$, $\delta=0.16$, $l=5.3$~nm,
 and $L=1.12\times 10^4$ for the Shaker
and 
$S_V=8.91 \times 10^{-3}$, $\delta=0.16$, $l=5.5$~nm,
 and $L=1.00\times 10^4$ for the MaxiK.
The parameters $s$ and $D$ have been computed via (\ref{esse}) and (\ref{di}).}
\label{t:nuova}
\end{table*}

Let $I$ be the measured current, 
$V$ the voltage applied across the cellular membrane
and write 
\begin{equation}
\label{iva}
I=
S_I f
\end{equation}
where $f$ is given in (\ref{flussok}) with 
\begin{equation}
\label{nelson}
v=
S_V
\frac{e^{\alpha \delta V}-e^{-\alpha \delta V}e^{-\alpha(1- 2\delta) V}}
     {1+e^{-\alpha(1- 2\delta) V}}
\end{equation}
where 
$S_I,S_V,\delta$ are positive parameters and $\alpha$ is the constant 
$\alpha=e_0/K_b T=0.039$~mV$^{-1}$, with 
$e_0=1.6\times10^{-7}$~pC the charge of the potassium ion, 
$K_\rr{b}$ the Boltzmann constant, and $T=298.15$~$^\rr{o}$K the temperature.

As we discussed in the Introduction, the choice of the probability $v$ 
as a function of the voltage $V$ is the only ingredient of the model 
related to the dynamics of the ions inside the filter. 
The choice (\ref{nelson}) is inspired by the model in 
\cite{Nelson,Nelson02,Nelson03} 
and dates back to the knock--on model in \cite{HK02}.
We remark that, as it will be made clear in Section~\ref{s:nelson}, the ability
of our model to describe the experimental results does not depend
strictly on this choice. See also the comments there in this connection.

We fit the experimental data for concentration $800$~mM
by using the above formula with $N=800$. 
In this way we fix the parameters
\begin{displaymath}
S_I,\; S_V,\; \delta,\; L, \textrm{ and } p_{800}
\end{displaymath}
where $p_{800}$ is the low--affinity probability 
for concentration $800$~mM. 

\begin{figure*}
\includegraphics[height=10.cm,angle=0]{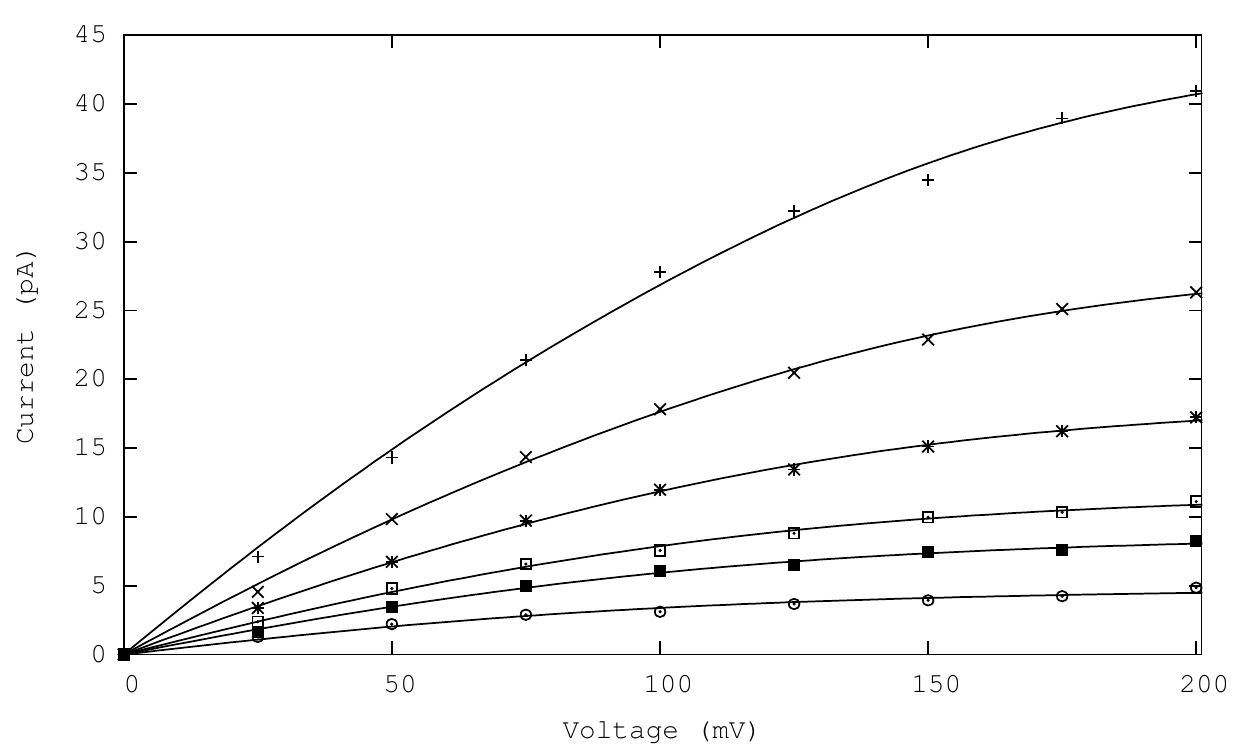}
\caption{Comparison between experimental KcsA 
potassium current--voltage curves in 
\cite{MHM} (symbols) and theoretical prediction (solid lines).
The symbols $\bigcirc$, $\blacksquare$, $\square$, 
$*$, $\times$, and $+$ 
refer respectively to the potassium concentrations
$20,50,100,200,400,800$~mM.}
\label{f:fit-k-a}
\end{figure*}

We complete the analysis of the data set according to 
the following scheme. 
The values of $S_I$ and $p$ for the six experimental series 
are supposed to change, that is to say we assume that 
the probability that the filter is in 
the low--affinity state and the constant $S_I$ (we will 
see in Section~\ref{s:discussione} that this constant is
related to the diffusion coefficient in the intracellular region)
depend on the ionic concentration. 
The five other experimental series, that is to say those 
measured at concentrations $400,200,100,50$, and $20$~mM, are then 
fitted by the same equation by keeping fixed 
$S_V$, $\delta$, $\alpha$, and $L$ and by using $p$ and $S_I$ as the sole 
fitting parameters. Results are plotted in figure~\ref{f:fit-k-a};
the fitted parameters are reported in table~\ref{t:nuova}.
The physical meaning of the fitting parameters will be discussed 
in Section~\ref{s:discussione}.

This fitting scheme is based on a simple remark: we assume
that the channel behavior, that in our description is modeled
by the function (\ref{nelson}), does not depend on the 
ion concentration. On the other hand we cannot exclude that 
the ionic diffusion coefficient in the intracellular region
\cite{HS}
and the typical 
time spent by the pore in the low--affinity state 
depend on the concentration of the ionic species in the 
cell.

The soundness of the values that we found for the fitting parameters
will be discussed in Section~\ref{s:discussione}. We now 
mention only two facts: to our knowledge there exists no 
experimental measure of the 
time fraction spent by the filter in the low--affinity state. 
So that we cannot say anything about the validity of our 
estimate for $p$; we can just remark the qualitative 
agreement with \cite[Figure~5]{VanDongen01}.
The estimated diffusion coefficient
is several orders of 
magnitude larger than the real (experimental) value \cite{HS,HK}. 
This expected problem is due to the one--dimensional 
character of the model. In Section~\ref{s:discussione} we shall 
compute the associated three--dimensional estimates that will 
result to be in very good agreement with the experimental values.

\begin{figure*}
\includegraphics[height=10cm,angle=0]{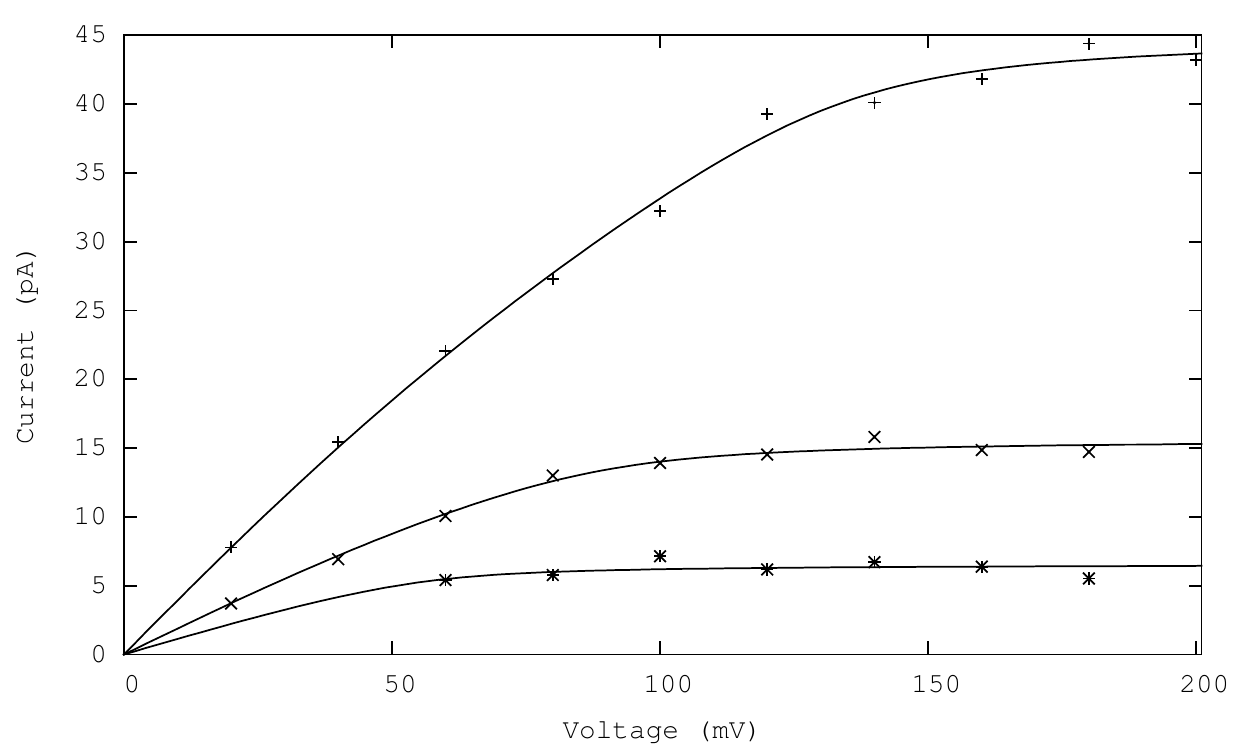}
\caption{Comparison between experimental MaxiK
potassium current--voltage curves in 
\cite{SH} (symbols) and theoretical prediction (solid lines).
The symbols 
$*$, $\times$, and $+$ 
refer respectively to the potassium concentrations
$50,150,400$~mM.}
\label{f:fit-SH}
\end{figure*}

Finally we note that
with our fitting scheme we also are able to reproduce with good accuracy
the data in \cite{SH}, and \cite{HmK} referring, respectively, to the 
MaxiK and Shaker channel in \cite{HmK} 
(see figure~\ref{f:fit-SH} and table~\ref{t:nuova}).
Data in figure~\ref{f:fit-SH} have been reproduced with 
the authors' permission.
The data related to the Shaker channel have been 
extracted from figure~1B in \cite{HmK}.

\subsection{Potassium conductance curves}
\label{s:conduttanza}
\begin{figure*}
\includegraphics[height=10cm,angle=0]{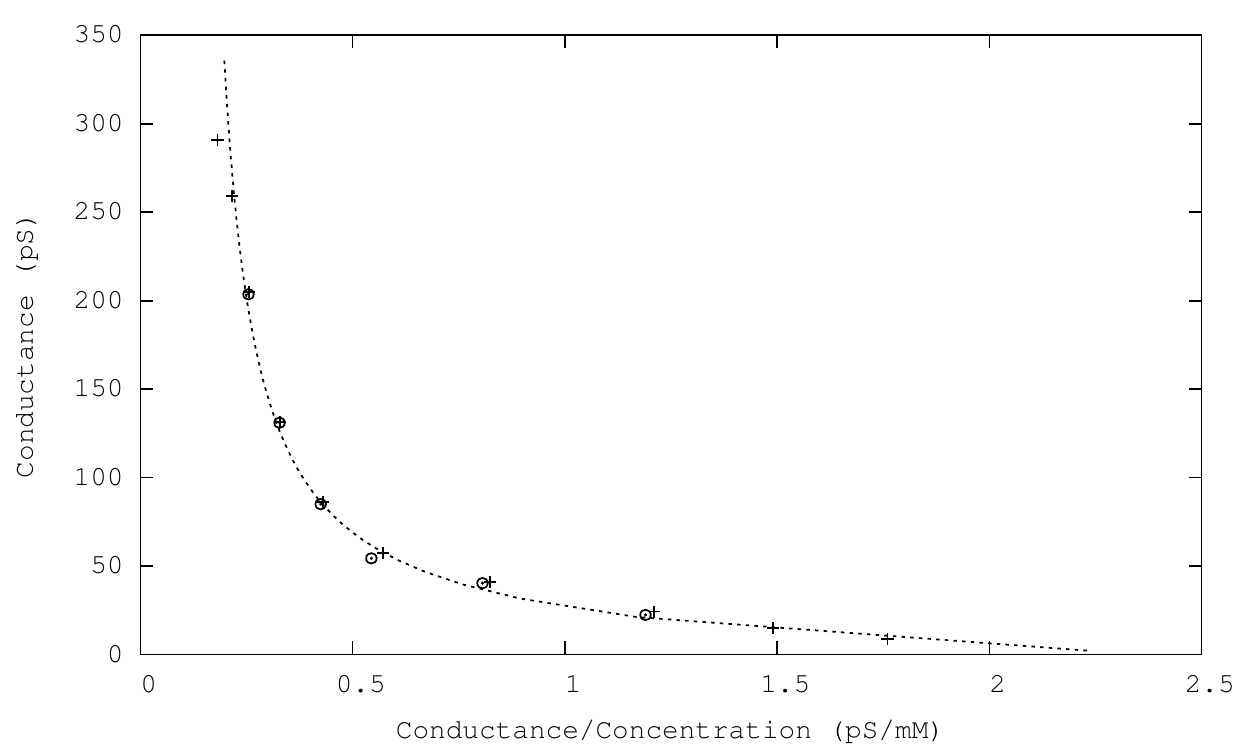}
\caption{The Eadie-Hofstee plot for the potassium permeation current 
at external voltage $200$~mV. 
The symbols $+$ and $\bigcirc$ refere, respectively, to 
the experimental data and to the theoretically computed values.
The dotted line is an eye guide.} 
\label{f:fit-cond-a}
\end{figure*}

In \cite{MHM} the permeation behavior of the pore 
has been investigated also by means of the conductance $g=I/V$.
We compared the permeation data in 
\cite{MHM} at fixed voltage $V=200$~mV and concentration 
varying from $5$~mM to $1600$~mM, with the results of 
our fit of potassium current--voltage curves 
(see Section~\ref{s:potassio}), as shown in 
figure~\ref{f:fit-cond-a} by an 
Eadie--Hofstee plot. 
In the picture
the symbols $+$ and $\bigcirc$ refer, respectively, to 
the experimental data and to the theoretically computed values, whereas
the dotted line is an eye guide. 

The matching is good in the whole region where full experimental data sets 
are known. This graph shows neatly the ability to our model 
to capture also the dependence of the permeation current 
on the ion intracellular concentration. 

Even if we do not have any physical argument to support this 
choice, it is worth noting that 
the dotted line in the picture, which is just an eye guide, 
has been indeed obtained by assuming a (slightly sub--linear) power low
dependence of the low--affinity state probability $p$ on the 
ion concentration and an Hill type behavior for the diffusion coefficient. 

\begin{table*}
\begin{center}
\begin{tabular}{c|c|c|c|}
\cline{2-4}
 & $NH4^{+}$ & $Tl^{+}$ & $Rb^{+}$ \\
\hline
\multicolumn{1}{|c|}{$\delta$} 
&$0.17$
&$0.017$
&$0.33$\\
\hline
\multicolumn{1}{|c|}{$S_V$} 
&$3.2\times 10^{-4}$  
&$6.8\times 10^{-4}$ 
&$9.4\times 10^{-5}$ \\
\hline
\end{tabular}
\end{center}
\caption{Parameters measured by fitting 
the experimental data for ions different from potassium 
in KcsA via (\ref{iva}) and (\ref{nelson}). Recall also that $N=100$,
$S_I=8.07\times 10^{6}$, $p=6.1\times 10^{-4}$, 
$s=1.98\times 10^{-14}$~sec, $D=7.62 \times 10^{-2}$~cm$^{2}$/sec 
and $L=10^4$.}
\label{t:fit-altri}
\end{table*}

\subsection{Other species current--voltage curves}
\label{s:altri}
\par\noindent
We test our model against the experimental
current--voltage curves \cite{MHM}
for ions different from potassium. 
More precisely
we consider the curves for NH$_4^{+}$, Tl$^{+}$, and Rb$^{+}$
at ionic concentration 100~mM.

To fit these curves we fix the values of $S_I$, $L$, and $p$ 
to the ones found previously for potassium at 
concentration 100~mM, and we use $\delta$ and $S_V$
as fitting parameters.
That is to say that we assume that the
ionic diffusion in the intracellular region and the probability 
of finding the filter in the low--affinity state are 
independent on the ionic species, but depend only on the 
ionic concentration. 
This is suggested by the measured values of the diffusion coefficient,
see for instance \cite[Table~2]{CZSO} for Tl$^+$ and 
\cite[Table~1.1-1]{Cussler} for the other ions. 

\begin{figure*}
\includegraphics[height=10.cm,angle=0]{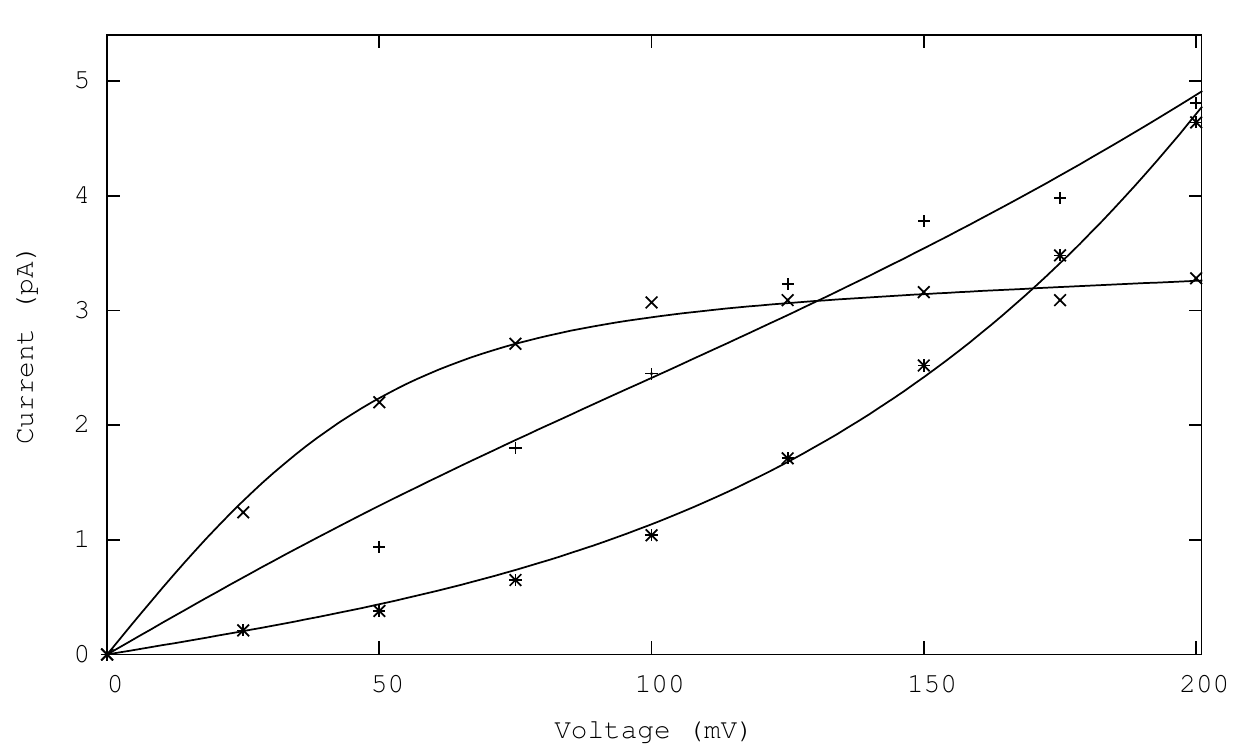}
\caption{Comparison between experimental current--voltage curves  
\cite{MHM} (symbols) and theoretical prediction (solid lines).
The symbols  $*$, $+$, and $\times$
refer respectively to Rb$^+$, NH$_4^+$, and Tl$^+$
at concentration $100$~mM.}
\label{f:fit-oth-a}
\end{figure*}

With this assumption we are implicitly saying that the 
reduced, with respect to potassium, 
permeation rates typical of the other selected ion species is 
due to the behavior of such ions inside the selectivity filter. 
Results are plotted in figure~\ref{f:fit-oth-a};
the fitted parameters are reported in table~\ref{t:fit-altri}. 

Deeper investigation on this point should be corroborated by 
additional data. As in the potassium case, data sets 
corresponding to different values of the intracellular concentrations 
are needed. 

\subsection{Modeling the channel}
\label{s:nelson}
\par\noindent
In the discussion above the choice of the relation between the 
ejection probability 
and the voltage across the membrane, see equation (\ref{nelson}), 
has been inspired by the physical argument proposed in the kinetic model 
for single--file 
ion channels in \cite{Nelson,Nelson02,Nelson03}. 
The function in (\ref{nelson}) has the same form 
of the channel permeation rate used in that model, 
that is to say we are assuming
that the inner part of the channel behaves like a single file ion channel.
One of the key novelties 
in our work is the coupling between this mechanism and
the diffusion of the ions in the intracellular region. 

It is worth noting that the analysis conducted 
in Subsections~\ref{s:potassio}, \ref{s:conduttanza}, and \ref{s:altri}
can be 
repeated using a simple power law (having no particular physical meaning)
for the ejection probability as a function of the external voltage.
In other words we can assume 
\begin{equation}
\label{ivb}
v=\Big(\frac{V}{C_V}\Big)^\gamma
\end{equation}
for modeling the relation between the ejection probability 
and the voltage across the membrane.

In this case we get similarly good results, that is to say 
we are able to reproduce 
the experimental data with good precision.
We do not show the graphs, which are similar to those in 
figure~\ref{f:fit-k-a}, but report the whole set of 
fitted parameters in table~\ref{t:potenza}.

This remark suggests that, in the framework of our model,
no particular modeling of the 
dynamics of ions inside the channel is needed in order to get 
the correct behavior of the current--voltage curves. 
But it is important to remark that 
different filter models produce different values for the 
time fraction spent by the selectivity filter in the low--affinity 
state. 
Indeed, see tables~\ref{t:nuova} and \ref{t:potenza},
by using the models (\ref{nelson}) and (\ref{ivb})
for the dynamics inside the filter, similar values for the 
diffusion coefficient are found, whereas the predictions for the 
time fraction $p$ spent by the selectivity filter in the low--affinity 
state differ by an order of magnitude. 
This suggests that the current--voltage curves 
could be a useful instrument to predict the  
time fraction spent by the selectivity filter in the low--affinity 
state once the dynamics of the ions in the filter is known
or vice--versa. 

\begin{table*}
\begin{center}
\begin{tabular}{c|c|c|c|c|c|c|}
\cline{2-7}
& 
\multicolumn{6}{|c|}{KcsA}
\\
\hline
\multicolumn{1}{|c|}{$N$} 
 & $800$ & $400$ & $200$ & $100$ & $50$ & $20$ 
\\
\hline
\multicolumn{1}{|c|}{$S_I$ ($10^6$~pA)} 
 &$4.47$ &$6.03$ &$7.71$ &$9.73$ &$14.5$ &$19.6$ 
\\
\hline
\multicolumn{1}{|c|}{$p$ ($10^{-2}$)} 
 &$3.02$ &$1.77$ &$1.14$ &$0.82$  &$0.45$ &$0.39$ 
\\
\hline
\hline
\multicolumn{1}{|c|}{$s$ ($10^{-14}$~sec)} 
 &$3.38$ &$2.65$ &$2.07$ &$1.64$  &$1.11$ &$0.82$ 
\\
\hline
\multicolumn{1}{|c|}{$D$ ($10^{-2}$~cm$^{2}$/sec)} 
 &$4.47$ &$5.70$ &$7.29$ &$9.20$  &$13.7$ &$18.5$ 
\\
\hline
\end{tabular}
\end{center}
\caption{The parameters $S_I$ and $p$ have been  
measured by fitting the experimental data in 
\cite{MHM} via (\ref{iva}) and (\ref{ivb}). 
By fitting the highest concentration set of data 
we also found 
$C_V=6.31\times 10^{4}$, $\gamma=1.19$, $l=5.5$~nm and $L=1.00\times 10^4$.
The parameters $s$ and $D$ have been computed via (\ref{esse}) and (\ref{di}).}
\label{t:potenza}
\end{table*}

\section{Discussion}
\label{s:discussione}
\par\noindent
In the above section we have seen that the model proposed in this 
paper is able to explain in great detail the experimental data 
for the current--voltage curves in the KcsA ionic channel. 
We have shown that the theoretical predictions
fit nicely the measured curves provided the parameters defining 
the model are suitably chosen. 
It is notable that our model explains the dependence
of the permeation current both on the external voltage 
(figures~\ref{f:fit-k-a}, \ref{f:fit-SH}, and \ref{f:fit-oth-a})
and on the ion concentration in the intracellular region 
(figure~\ref{f:fit-cond-a}).

In \cite{Nelson02} the same experimental 
data \cite{MHM} have been analyzed via a kinetic model formerly 
introduced by the same author. 
In that model the selectivity filter is treated as an isolated structure
and the intracellular ion concentration is an input constant of the model.
Two different regimes 
for the filter had to be assumed corresponding, respectively, to high 
(400~mM and 800~mM) and low (20~mM, 50~mM, and 100~mM) 
potassium concentrations. 
Two different sets of the parameters carachterizing the filter
behavior (the analogue of $S_V$ and $\delta$) were found.

In other words the model in \cite{Nelson02} 
predicts that the behavior of the selectivity 
filter depends on the intracellular ion concentration. 
In our model the selectivity filter is coupled 
with the diffusion of ions in the intracellular region; this is 
indeed the key novelty and the distinguishing feature of our approach.
This allows to explain the experimental 
data referring to all the concentrations 
with the same set of parameters, $S_V$ and $\delta$, 
for the filter behavior. 

To compare the way in which data are explained by the model in 
\cite{Nelson02} and by our model, we first note that 
the electrical dissociation distance $\delta$ 
(see \cite[Fig.\ 1]{Nelson}) that we 
find is very close to that fitted there.
Moreover, we note that in \cite{Nelson02} in the lower intracellular 
concentration regime, the fitted parameters are such that an higher 
effective concentration ($c^*$ in their notation) is seen. 
In our model particle diffusion accounts 
for this effect. 

We now discuss 
the reasonableness of the values we found for these fitting parameters
for the potassium current data for the KcsA ionic channel 
(see Section~\ref{s:potassio}).

As a first step we have to give a continuum interpretation of
the lattice modeling the cytosol. In other words we have to associate 
with the unit length of the lattice model 
a physically reasonable quantity. 
We imagine to associate with each lattice site a small cubic volume 
whose side length is denoted by $\ell$.  
Recall that the experimental concentration $c$ is 
expressed in millimolar, that is as number of moles per cubic meter 
(i.e., number of millimoles per liter). Recall, also, that $N$ is the number 
of particles in the lattice model and that $L$ is the number of 
sites in $\Lambda$. We then have the following identification
\begin{displaymath}
c=\frac{N}{N_\rr{A}} \frac{1}{\ell^3 L}
\end{displaymath}
where $N_\rr{A}=6.022\times10^{23}$ is the Avogadro number and 
$\ell$ has to be expressed in meter.

Since in our fitting procedure we have chosen $N$ such that 
$c/N=1$~mM, we have that 
\begin{equation}
\label{elle}
\ell=\frac{1}{(N_\rr{A}L)^{1/3}}
=
5.50\times10^{-10}~\textrm{m}=0.55~\textrm{nm}
\end{equation}
where we have used that $L$ has been fitted to 
$10^4$, see the caption of table~\ref{t:nuova}.
The above argument provides us with a way to associate 
a length unit measure with our lattice model. The result we found for $\ell$
is physically reasonable, indeed we can assume that each 
cubic volume associated with a site of the lattice can accommodate 
few (say one or two) ions, so that 
its side length has to be of the order of magnitude
of the ionic diameter. 
This value we found for $\ell$ is then reasonable, since 
potassium 
ion, atomic, and van der Waals radius are respectively given by 
$0.13, 0.23$, and $0.27$~nm.

As a second step we have to give a continuum interpretation of
the discrete unit of time of the Markov Chain. 
In other words we have to associate 
with the unit of time a physically reasonable quantity $s$. 
We consider the evolution of our model up to the time $n$ and  
note that 
the true current measured in the experiments 
can be identified as 
\begin{displaymath}
I
=
\frac{e_0\times\textrm{(number of particles which exited }\Lambda\textrm{)}}
      {s\times\textrm{(number of iterations)}}
=
\frac{e_0}{s}\,f
\end{displaymath}
where 
$e_0=1.6\times10^{-7}$~pC 
is the charge of a potassium ion and 
$f$ is the theoretical model outgoing flux. 
We then identify 
$e_0/s$ with the parameter $S_I$ introduced in equation (\ref{iva});
that is to say we write 
\begin{equation}
\label{esse}
s=\frac{e_0}{S_I}
\end{equation}
and we
can use the estimated $S_I$ listed in the first row of  
table~\ref{t:nuova} to compute the values of $s$ listed in the 
same table.

We have no direct clue to establish if 
these values for $s$ are reasonable or not. But starting from this values 
we can estimate the diffusion coefficient of potassium ions in 
the cytosol. 
Indeed, see Appendix~\ref{s:diff}, the intracellular ion diffusion coefficient 
is related to the other parameters by the equation 
(\ref{ident}), which yields
\begin{equation}
\label{di}
D=\frac{\ell^2}{2s}
\end{equation}
By using this equation we compute the diffusion coefficient
for the potassium. Results have been reported in the last row 
of table~\ref{t:nuova}.

The order of magnitude we found for the diffusion coefficient 
of potassium ions 
is $10^{-2}$~cm$^2/$sec. 
This result is several (say three) order of 
magnitude larger than the real (experimental) value \cite{HK}. 
This problem was indeed quite expected, since we modeled the 
diffusion of ions in cytosol with a one--dimensional lattice. 
As we have already remarked, the one--dimensional choice 
is motivated by the possibility to write explicitly the solution 
of the probabilistic model. Explicit formulas are quite necessary 
to perform the extended analysis discussed in 
Section~\ref{s:sperimentali}. As a future work 
we are now planning a 
Monte Carlo assisted analysis for the three--dimensional 
version of our model.

In any case we prove, now, that the estimates we found in this 
one--dimensional case for the diffusion coefficient are quite reasonable.
Consider a simple symmetric random walk with unitary time 
on a three-dimensional cubic lattice 
with $L$ sites and spacing $m$. 
We consider the side length $L^{1/3}$ 
in order to ensure that 
the ``real" volume associated with the three dimensional model
is equal to that associated with the one dimensional one. 
By a classical argument similar to the one developed 
in the Appendix~\ref{s:diff}
it is proven 
that the \textit{squared mean distance} walked up to time 
$t$ is $6D_3t$ where $D_3$ is the diffusion coefficient. 
This implies that the typical time needed by the walker, started 
at random in the lattice, to reach a particular point of the 
boundary (say the pore) is of order
\begin{displaymath}
\frac{(mL^{1/3})^2}{64}
\times
\frac{1}{6D_3}
\times 6(L^{1/3})^2
\end{displaymath}
Indeed, 
the first term is the square of the average distance of a point 
inside a cube of side length $mL^{1/3}$
from the boundary of the cube itself 
and the third one is 
the number of times the walker has to visit 
the boundary before touching the pore (number of points on the 
lateral surface of the cube). 
Note that the product between the first and the second term is an estimate for 
the time needed to touch the boundary 
of the cube for a walker started at random in the cube itself.

In our former analysis, in computing fluxes, we indeed evaluated this time 
in the framework of our one--dimensional model. 
By repeating the same argument we can say that this time was estimated 
as
\begin{displaymath}
\frac{1}{16}(mL)^2
\times
\frac{1}{2D_1}
\times 2
\end{displaymath}
By equating the two expressions found above we find 
\begin{displaymath}
D_3=\frac{1}{4}D_1L^{-2/3}\approx 0.00054\times D_1
\end{displaymath}
where we used $L=10^4$ (see the caption of table~\ref{t:nuova}).
By the equation above it follows that to 
the diffusion constant estimated via the one--dimensional model 
ranging in $3.62$ -- $15.8\times10^{-2}$~cm$^2$/sec 
(see table~\ref{t:nuova})
it corresponds a ``true" diffusion coefficient 
ranging in $1.95$ -- $8.53\times10^{-5}$~cm$^2$/sec, 
which is very close to the experimental value for the 
potassium diffusion coefficient \cite{HK}. 

\section{Conclusions}
\label{s:conclusioni}
\par\noindent
In this paper we have 
studied the effect of intracellular ion diffusion on 
ionic currents permeating through the cell membrane ionic channels.
Ion channels
share the following common properties: the ion flux is rapid, 
the channel is selective, and its functions are regulated 
by a gating mechanism. 
Although a formidable effort, with absolutely remarkable results, 
has been performed recently, 
a complete understanding of 
ionic channel behavior is still lacking. 

Thanks to the patch clamp technique
very precise measurements of  
the true current across a single channels are 
available. 
In this paper we have proposed a model 
which is able to provide a full 
explanation of the KscA current--voltage experimental 
curves \cite{MHM} by taking into account the following three 
effects: 
ion diffusion in the intracellular region, 
dynamics of the filter, 
and dynamics of the ions inside the filter.  

In particular our model points out the role played 
by intracellular diffusion on ion permeation in compensating 
the ion depletion in the region close to the pore. 
The model is able to explain the dependence of the 
experimetal data referring to the true ionic currents 
both on the external voltage and on the intracellular potassium 
concentration. 

Is is also notable that all the fitting parameters have a clear 
qualitative and quantitative physical interpretation. 
Moreover, we noted that the estimate we got for the typical time 
spent by the filter in the low--affinity state is strictly 
related to the dynamics of ions inside the filter. 
This suggests that our approach, corroborated by new experimental 
data, should be able to shed some light also on the dynamics 
of the ions inside the filter. 

\appendix
\section{Detailed definition of the model}
\label{s:bold}
\par\noindent
We consider an integer \textit{time} variable $n$. We set 
$n=0$ and choose at random with uniform probability $1/L$ 
the position of the $N$ particles.
We then repeat the following \textit{steps}:
\begin{enumerate}
\item
set $n$ equal $n+1$;
\item
select at random the state of the pore: 
choose the low--affinity state with probability 
$\plow$ and the high--affinity one with probability
$1-\plow$;
\begin{enumerate}
\item
if the pore is in the low--affinity state and it is 
occupied by a particle,
the particle is released with the following rule:
it jumps with probability $1-v$ to the site of 
$\Lambda$ neighboring the pore or, with probability $v$,
it exits the system;
\item
if a particle exits the system, a particle of the same species is 
put at random with uniform probability $1/L$ on one of the $L$ sites 
in $\Lambda$;
\end{enumerate}
\item
the position of each particle on the lattice is updated 
following the rules defined in Section~\ref{s:modello};
\begin{enumerate}
\item
if a particle enters the pore and the pore is in the 
low--affinity state, the particle is immediately released by the pore 
with the following rule:
it jumps with probability $1-v$ to the site of 
$\Lambda$ neighboring the pore or, with probability $v$,
it exits the system;
\item
if a particle exits the system, a particle of the same species is 
put at random with uniform probability $1/L$ on one of the $L$ sites 
in $\Lambda$.
\end{enumerate}
\end{enumerate}

\section{Diffusion coefficient}
\label{s:diff}
\par\noindent
Consider a one--dimensional walker on $\ell\bb{Z}$ and 
denote its position at time $t\in s\bb{Z}_+$ by
\begin{displaymath}
S_t\in\{\dots,-\ell,0,+\ell,\dots\}
\end{displaymath}
The random variable 
$S_t$ is the sum of $t/s$ (note that $t/s$ is a positive integer)
independent identically distributed
random variables $Y_i\in\{-\ell,+\ell\}$
such that 
\begin{displaymath}
\bb{P}(Y_i=+\ell)=
\bb{P}(Y_i=-\ell)=\frac{1}{2}
\end{displaymath}
A straightforward computation
yields
\begin{displaymath}
\bb{E}[Y_i]=0
\textrm{ and }
\bb{E}[(Y_i-\bb{E}[Y_i])^2]=\ell^2
\end{displaymath}
for the average and the variance of the 
random variable $Y_i$ respectively.
By independence we also get
\begin{displaymath}
\bb{E}[S_{t/s}]=0
\textrm{ and }
\bb{E}[(S_{t/s}-\bb{E}[S_{t/s}])^2]=\frac{t}{s}\ell^2
\end{displaymath}

By the Central Limit Theorem 
(see, for instance, \cite[Theorem (4) in Section~5.10]{GS}) 
we have that for $t$ large 
the probability density of the random variable $S_{t/s}$ is very 
well approximated by 
the Gaussian
\begin{displaymath}
\frac{1}{\sqrt{2\pi\ell^2(t/s)}} \exp\{-x^2/[2\ell^2(t/s)]\}
\end{displaymath}
for $x\in\bb{R}$.
By comparing this result with the continuum description given 
by a one--dimensional diffusion equation $\dot{u}=Du''$ 
with diffusion coefficient $D$
we get the identification
\begin{equation}
\label{ident}
2Ds=\ell^2
\end{equation}
allowing to compare lattice and continuum results.

\begin{acknowledgements}
The authors thank I.\ Schroeder and U.P.\ Hansen for having 
provided the experimental data in figure~\ref{f:fit-SH}.
The authors thank I.\ Chiarotto and M.\ Feroci 
for having suggested some of the references. 
\end{acknowledgements}

\newpage

\newpage


\newpage

\newpage

\newpage


\begin{thebibliography}{10}
\bibitem{HH}
A.L.~Hodgkin and A.F.~Huxley,
J.\ Physiol.\ \textbf{116}, 473 (1952).

\bibitem{NS}
E.~Neher and B.~Sakmann,
Nature \textbf{260}, 799 (1976).

\bibitem{Hille}
B.\ Hille,
``Ion Channels of Excitable Membranes," Third Edition,
Sinauer Associates Inc., Sunderland, MA, Usa, 2001.

\bibitem{GBOZ}
S.A.N.~Goldstein, D.~Bockenhauer, I.~O'Kelly, and N.~Zilberberg,
Nature Reviews Neuroscience \textbf{2}, 175 (2001).

\bibitem{VanDongen02}
A.M.J.~VanDongen,
Comm.\ Theor.\ Biol.\ \textbf{2}, 429 (1992).

\bibitem{Miller}
C.~Miller, 
Genome Biology \textbf{1} reviews0004 (2000).

\bibitem{FH}
D.\ Fedida and J.C.\ Hesketh,
Prog.\ Bio.\ Mol.\ Biology \textbf{75}, 165 (2001).

\bibitem{RCM}
M.\ Recanatini, A.\ Cavalli, and M.\ Masetti,
ChemMedChem \textbf{3}, 523 (2008).

\bibitem{ADSEGHTM}
A.~Abenavoli, M.L.~Di Francesco, I.~Schroeder,
S.~Epimoshko, S.~Gazzarini, U.P.~Hansen, G.~Thiel, and A.~Moroni,
J.\ Gen.\ Physiol.\ \textbf{134}, 219 (2009).

\bibitem{ZM}
Y.\ Zhou and R.\ Mackinnon,
J.\ Mol.\ Biol.\ \textbf{333}, 965 (2003).

\bibitem{doyle}
D.A.\ Doyle, C.J.\ Morais, R.A.\ Pfuetzner, A.\ kuo, 
J.M.\ Gulbis, S.L.\ Cohen, B.T.\ Chait, R.\ MacKinnon,
Science \textbf{280}, 69--77 (1998).

\bibitem{IOTS}
S.\ Imai, M.\ Osawa, K.\ Takeichi, I.\ Shimada,
PNAS \textbf{107}, 6216--6221 (2010).


\bibitem{Aqvist}
J.\ \r{A}qvist and V.\ Luzhkov,
Nature \textbf{404}, 881 (2000).

\bibitem{BR}
S.\ Bern\`eche, B.\ Roux,
PNAS \textbf{100}, 8644--8648 (2003).

\bibitem{Miller02}
C.\ Miller,
J.\ Gen.\ Physiol.\ \textbf{113}, 783 (1999).

\bibitem{Nelson}
P.H.\ Nelson,
J.\ Chem.\ Phys.\ \textbf{117}, 11396 (2002).

\bibitem{Nelson02}
P.H.\ Nelson,
Phys.\ Rev.\ E \textbf{68}, 061908 (2003).

\bibitem{Nelson03}
P.H.\ Nelson,
J.\ Chem.\ Phys.\ \textbf{134}, 165102 (2011).

\bibitem{MP}
S.\ Maf\'e and J.\ Pellicer,
Phy.\ Rev.\ E \textbf{71}, 022901 (2005).

\bibitem{VanDongen01}
A.M.J.~VanDongen,
PNAS \textbf{101}, 3248 (2004).

\bibitem{MHM}
M.\ LeMasurier, L.\ Heginbotham, and C.\ Miller
J.\ Gen.\ Physiol.\ \textbf{118}, 303--313 (2001).

\bibitem{SH}
I.\ Schroeder, U.P.\ Hansen, 
J.\ Gen.\ Physiol.\ \textbf{130}(1), 83 (2007).

\bibitem{HmK}
L.\ Heginbotham, R.\ MacKinnon, 
Biophysical Journal \textbf{65}, 2089--2096 (1993).

\bibitem{ABCM01}
D.\ Andreucci, D.\ Bellaveglia, E.N.M.\ Cirillo, S.\ Marconi, 
Physical Review E \textbf{84}, 021920 (2011).

\bibitem{RMFMPEAFGMG}
M.L.\ Renart, E.\ Montoya, A.M.\ Fern\'andez, M.L.\ Molina, J.A.\ Poveda,
J.A.\ Encinar, J.L.\ Ayala, A.V.\ Ferrer--Montiel, J.\ G\'omez, A.\ Morales,
and J.M.\ Gonz\'alez Ros,
Biochemistry \textbf{51}, 3891--3900 (2012).

\bibitem{HK02}
A.L.~Hodgkin and R.D.~Keynes,
J.\ Physiol.\ \textbf{128}, 61 (1955).

\bibitem{HS}
H.S.\ Harned and J.A.\ Shropshire,
J.\ of the American Chemical Society \textbf{80}, 5652--5653 (1958).

\bibitem{HK}
A.L.~Hodgkin and R.D.~Keynes,
J.\ Physiol.\ \textbf{119}, 513--528 (1953).

\bibitem{CZSO}
M.\ Ciszkowska, L.\ Zeng, E.O.\ Stejskal, and J.G.\ Osteryoung
J.\ Phys.\ Chem.\ \textbf{99}, 11764--11769 (1995).

\bibitem{Cussler}
E.L.\ Cussler,
``Diffusion."
Second Edition, Cambridge University Press, Cambridge, UK, 1997.

\bibitem{GS}
G.\ Grimmet, D.\ Stirzaker,
``Probability and Random Processes."
Oxford University Press Inc., New York, US, 2001. 




\end{thebibliography}
\end{document}